\documentstyle[editedvolume,psfig,namedreferences]{crckapb}
\begin{opening}
\title{Future Possibilities for Detecting HI at High Red-shift}
\author{Robert Braun}
\institute{Netherlands Foundation for Research in Astronomy} \end{opening}
\runningtitle{Detecting HI at High Red-shift}
\begin{document}
\vskip -1cm

\centerline{\it To appear in: ``Cold Gas at High Redshift''}
\centerline{\it Eds. M.Bremer et al. (Kluwer, Dordrecht)}

\section{Introduction}

The 21-cm line of neutral hydrogen is in many respects the most
valuable tracer of neutral gaseous mass in astrophysics. Even though
neutral gas becomes predominantly molecular at high densities, the
21-cm line emission of the associated atomic component allows the total
gaseous mass of (proto) galactic concentrations to be estimated to
about a factor of two, even in the most extreme cases
observed to date. This is in marked contrast to, for example, the
luminosity of carbon monoxide emission lines originating in the
molecular component. For the same total gaseous mass, these emission
lines are observed to vary in luminosity over a factor of about 10$^4$
depending on the abundance of heavy elements and the intensity of
illumination to which they are subjected.

It is also important to draw the distinction between the sensitivity to
the neutral gaseous mass of distinct concentrations with the
sensitivity to column density along discrete lines-of-sight. The
extremely large absorption cross-section of neutral hydrogen to
Lyman-$\alpha$ radiation guarantees a correspondingly high sensitivity
to neutral hydrogen columns along lines-of-sight to suitable background
sources of (red-shifted) ultra-violet light. However, the paucity and
compact nature of such background probes implies that gas masses can
never be determined for individual objects via such observations of
absorption. An emission line tracer of the neutral gas remains
essential for this purpose.

In this paper we will briefly review the beginnings of HI astronomy,
proceed to an assessment of our current capabilities in this area, and
continue by considering what will be necessary to push back the
frontier to cosmological distances. We will then consider how such a
leap in performance might be realized.

\section{Where have we come from?}

After the prediction of the existence of the HI 21-cm line by
Van~de~Hulst (1945) it was only a few years before several independent
groups detected the line emission from our Galaxy and reported it
simultaneously in a 1951 issue of {\it Nature\ } (Ewen and Purcell,
Muller and Oort). Even in the preliminary discovery report,
important contributions were made by Muller and Oort (1951) to our
knowledge of the kinematics and morphology of the gaseous disk of our
Galaxy. This was followed, in short order, with major surveys of spiral
structure and kinematics of the Galaxy (Van de Hulst, Muller and Oort,
1954)

\section{Where are we now?}

The years since 1951 have seen a major improvement in our abilities to
observe HI in galaxies. An illustration of this improvement is given in
Figure~1 where the continuum sensitivity is plotted after one minute of
integration for many radio telescopes as they became available. Since
upgraded receiver systems have often been added to existing facilities
after construction, these are also indicated in the figure in a number
of cases. An exponential improvement in sensitivity over at least 6
orders of magnitude is apparent between about 1940 and 1980.
Instruments like the WSRT and the VLA have become available on a
schedule which maintained a high rate of discovery. The Arecibo
telescope stands out as a major leap in sensitivity performance at a
relatively early date.

\begin{figure}[htb]
\caption{
The time evolution of radio telescope sensitivity. The continuum
sensitivity after one minute of integration is indicated for a number of
radio telescopes as they became available. Solid lines indicate up-grade
paths of particular instruments. }
\end{figure}

Sensitivity within the narrow bandwidth of a spectral line observation
has evolved in a very similar way to that illustrated in Figure~1. We
have now progressed to the point where we can study the neutral gas
content, distribution and kinematics within individual galaxies at
recession velocities as high as about 24,000 km~s$^{-1}$. This is
perhaps best illustrated by a recent deep integration (about 100 hours)
with the VLA obtained by Van Gorkom (1995) to image HI
in galaxies of the cD cluster A2670. Some 20 cluster galaxies are
detected within an area of about 0.25 square degrees with an rms
sensitivity of 80~$\mu$Jy~beam$^{-1}$ per 48~km~s$^{-1}$ channel. The
corresponding 5$\sigma$ detection limit for a galaxy spanning
100~km~s$^{-1}$ is about $4\times10^8$h$^{-2}$~M$_\odot$.

As impressive as this result is, it also underlines the sad fact that
current instrumentation only allows us to determine neutral gas masses
for galaxian concentrations out to a red-shift of about 0.1. Given all
the evidence for substantial evolution of the gas mass at red-shifts
between perhaps 0.2 and 0.5, let alone the dramatic evolution expected
between z~=~0.5 and 2, this is particularly frustrating.

\section{Where are we headed?}

An overview of how current and upcoming instrumentation measures up to
the problem of detecting HI emission from distant systems is given in
Figure~2. The continuum and line emission of the luminous spiral galaxy
M101 has simply been re-scaled to simulate its appearance at the
indicated red-shifts of 0.12, 0.25, 0.5, 1, 2 and 4. For this
illustration, no time evolution of the emission spectrum has been
assumed, even though it is clear that the large stellar mass which is
now present was once also in the form of gas. The rms sensitivities of
a variety of existing and planned instruments (assuming a spectral
resolution of 10$^4$ and an integration time of 12 hours) have been
overlaid on these spectra. The HI emission line of such a gas-rich
system (M$_{HI}$=$2\times10^{10}$~M$_\odot$) is easily detectable by
the VLA and WSRT near z~=~0.1, but will probably demand the GMRT for
detection at z~=~0.3.

\begin{figure}[htb]
\caption{Red-shifted unevolved spectra of M101 compared to
the spectral line sensitivity of current and planned facilities at
frequencies between 10$^8$ and 10$^{12}$~Hz. The solid lines are the
composite emission spectra red-shifted to z~=~0.12, 0.25, 0.5, 1, 2  and
4.
Note that the continuum and spectral line features of the composite
spectra are sampled with different effective bandwidths.
Telescope names are placed at the rms sensitivity level after one
``transit'' of integration at a spectral resolution of 10$^4$. }
\end{figure}

An important point to note is that it is not merely a question of
having enough sensitivity to detect the HI emission line, but also that
the appropriate frequency coverage be available. For example, the VLA
20-cm band extends from 1320--1700~MHz at 0.9 times nominal
sensitivity, reaching only to z~=~0.08 in HI. The highest frequency band
of the GMRT on the other hand is expected to extend from 1000--1420~MHz,
so that red-shifts as high as 0.4 may become accessible.

Finally, it is essential that the frequency in question is not rendered
unusable by radio frequency interference. This is in fact the reason
that only synthesis arrays have been plotted in Figure~2. Experience has
shown that total power instruments, like Arecibo and the Green Bank 140
foot telescope are unable to achieve noise limited performance in those
portions of the spectrum which are in active use. For several reasons,
synthesis arrays are much less vulnerable to external interference. This
is an issue to which we will return below.

Another way of illustrating upcoming performance is given in Figure~3.
In this case the performance of the upgraded WSRT (as expected in 1997)
is illustrated for a long integration of 400~hr duration. The limiting
HI mass is plotted as function of red-shift for both the case of
``detection'' and ``imaging''. ``Detection'' is defined here as
requiring a 5$\sigma$ signal in a single 50~km~s$^{-1}$ velocity
channel, while ``imaging'' is defined as requiring a 5$\sigma$ signal
in each of six independent velocity channels of 50~km~s$^{-1}$ width.
The solid curves, which extend from z~=~0--0.2 and z~=~2.6--4.7,
indicate the frequency ranges where optimized receiver systems will be
available. Almost continuous coverage of the remaining interval,
z~=~0.2--2.6, will also be available for the first time with the new
receiver system, although at a reduced sensitivity.

\begin{figure}[htb]
\caption{
Current and future capabilities for detecting and imaging HI as a
function of red-shift for an integration of 400 hour duration.
``Detection'' is defined to imply a 5$\sigma$ signal in a single
50~km~s$^{-1}$ channel, while ``Imaging'' implies a 5$\sigma$ signal in
each of 6 independent channels of 50~km~s$^{-1}$ width. }
\end{figure}

The preceding discussion has illustrated how inadequate the current
generation of instruments will be to study galaxian gas masses out to
cosmological distances. This fact has been one of the major drivers for
pursuing a next generation facility with about two orders of magnitude
greater sensitivity than what is now available. Since the requirement
is basically for about 10$^6$~m$^2$ of collecting area, the proposed
facility has come to be called the ``Square Kilometer Array
Interferometer''.

The astute reader will already have noted that the capabilities of such
a new instrument have been overlaid on Figures~1, 2 and 3. From
Figure~1 it is clear that the SKAI sensitivity is what is required to
maintain the exponential improvement in performance that accompanied
the decades of broad ranging discovery between 1950 and 1990. Returning
specifically to the detection of the HI emission line, Figure~2
illustrates how individual galaxies, like M101, would be within the
reach of such an instrument out to red-shifts greater than 2. This
capability is placed in a more continuous context in Figure~3, where it
can be seen that SKAI will effectively open much of the universe to
direct study of (sub-)galaxian neutral gaseous masses and how they have
evolved.

While sufficient sensitivity to detect the integrated signal from
gaseous concentrations tells part of the story, another important
concern is having sufficient spatial resolution to allow kinematic and
morphological studies to be undertaken. The combined imaging and
detection capabilities of SKAI are illustrated in Figure~4, assuming
that most of the instrumental collecting area is concentrated in a
circular region of about 50~km in diameter. The angular resolution in
the red-shifted HI line then varies between about an arcsec locally to
2 arcsec by z~=~1. This combination of collecting area and array size
has been chosen to provide about 1~Kelvin of brightness sensitivity for
spectral imaging applications within a 24 hour integration. An actual
HI data-cube of M101 has been resampled and rescaled to simulate it's
appearance at the indicated red-shifts of 0.2, 0.45 and 0.9. The peak
observed brightnesses are shown in the left hand panels, while the
derived velocity fields are shown on the right. From the figure it is
clear that fairly detailed kinematic studies (including kinematic
detection of spiral arms, rotation curves, etc.) of ``normal'' systems
will be possible to at least z~=~0.5, while crude kinematics (basic
orientation and rotation parameters) will be possible to z~=~1 or more.
In addition, it should be borne in mind that the actual field-of-view and
spectral bandpass of a SKAI observation will be many times that shown in
Figure~4. While each panel of the figure is only 230~kpc on a side, the
likely SKAI field-of-view will correspond to about 1.5, 3 and 5~Mpc at
the three red-shifts shown. At the same time a total observing bandwidth of
100--200~MHz will probe a cylindrical volume about 200~Mpc deep. Each
pointed observation will therefore provide serendipitous kinematic
data on several hundred field galaxies.

\begin{figure}[htb]
\caption{
Simulated SKAI observations of M101 as it would appear at red-shifts of
0.2, 0.45 and 0.9. Peak observed brightnesses are shown in the left hand
panels and corresponding velocity fields on the right. The assumed
integration time is indicated above each panel. }
\end{figure}

So far we have assumed that the gas properties of galaxies at earlier
epochs are similar to those of current galaxies in making some
predictions of what might be achieved. It would be very surprising if
the universe were to behave in such a boring fashion. At the current
epoch, the vast majority of the gas that is gravitationally bound by
individual galaxies has already been cycled through, and to a great
extent locked in the form of, stars. If we can reach back to the time
when much of the early activity was taking place, which may correspond
to the quasar epoch between z~=~2--3, the current proportions may well
be reversed.

A better indication of what we might expect to find in the early
universe is beginning to emerge from extensive numerical simulations of
structure and galaxy formation (eg. Weinberg 1995, Ingram 1995). In
Figures~5, 6 and 7 we have taken the simulated neutral hydrogen densities
predicted by these simulations at red-shifts of 2, 3 and 4 (CDM
with $\Omega$~=~1, $\Omega_B$~=0.05, H$_0$~=~50~km~s$^{-1}$,
$\sigma_{16}$~=~0.7) in a co-moving volume that is 22.22/(1+z) Mpc on a
side and overlaid the five sigma detection contour of SKAI after a long
integration time of 1600 hr. Within a single pointing of the SKAI, and
each 2.5 MHz of spectral bandwidth we might expect a handful of
detections at z~=~4, perhaps 50 at z~=~3 and some hundreds at z~=~2.
Rather than being carried out as separate experiments, the instrumental
bandwidth and spectral resolution are likely to be sufficient to
observe the entire red-shift interval 2--4 simultaneously. Based on the
detection frequencies noted above, we would then expect such a single
experiment to allow study of some 8000 high red-shift systems.

\begin{figure}[htb]
\caption{
Simulated HI emission at z = 2 with SKAI detections overlaid. The
linear grey-scale indicates the predicted peak brightness of HI
emission in a 22.2/(1+z) Mpc cube and extends from $log(M_\odot/Beam)$
= 1.7 -- 10.8. The single white contour at $log(M_\odot/Beam)$ = 9.22
is the 5$\sigma$ SKAI detection level after a 1600 hour integration. }
\end{figure}

\begin{figure}[htb]
\caption{
Simulated HI emission at z = 3 with SKAI detections overlaid. The
linear grey-scale indicates the predicted peak brightness of HI
emission in a 22.2/(1+z) Mpc cube and extends from $log(M_\odot/Beam)$
= 2.6 -- 10.7. The single white contour at $log(M_\odot/Beam)$ = 9.72
is the 5$\sigma$ SKAI detection level after a 1600 hour integration. }
\end{figure}

\begin{figure}[htb]
\caption{
Simulated HI emission at z = 4 with SKAI detections overlaid. The
linear grey-scale indicates the predicted peak brightness of HI
emission in a 22.2/(1+z) Mpc cube and extends from $log(M_\odot/Beam)$
= 3.4 -- 10.7. The single white contour at $log(M_\odot/Beam)$ = 10.13
is the 5$\sigma$ SKAI detection level after a 1600 hour integration. }
\end{figure}

Of course the actual number and distribution of detections at these
red-shifts will probably be quite different than illustrated in
Figs.~5-7. However, those differences are likely to make it possible to
determine the cosmological model that actually applies to our universe.

\section{How will we get there?}

Some of the basic instrumental parameters of the SKAI have already
emerg\-ed from the previous discussion. The highest possible sensitivity
(corresponding to a baseline geometric collecting area of 10$^6$~m$^2$)
is required over frequencies from about 200--1400~MHz. This should be
coupled with the highest angular resolution that retains sufficient
brightness sensitivity for HI emission line detection. In practise this
implies an array distributed over a region of 30--50~km diameter. The
instantaneous field-of-view should be as large as possible (from
scientific considerations) while not limiting system performance on
long integrations at these relatively low frequencies. Considering both
ionospheric non-isoplanicity as well as sky model complexity suggests a
unit telescope size of between 100 and 300 meter diameter.
Instantaneous synthesized image quality must be sufficient to allow
adequate modeling of a time variable sky model (including ground and
space-based interfering sources), leading to a minimum requirement of
about 32 well-distributed units which would be cross-correlated.

The above requirements are embodied in the schematic
configuration shown in Figure~8. A densely packed elliptical zone
accounts for some 80\% of the array collecting area. The remaining 20\%
of the collecting area is distributed over a much larger region to
permit sub-arcsec resolution to be employed for other applications like
imaging in continuum radiation and HI absorption.

\begin{figure}[htb]
\caption{
Schematic configuration of SKAI. Note that the unit telescopes are not
depicted to scale. }
\end{figure}

Although the basic parameters of the instrument and its schematic
configuration can be derived in a straight forward manner, the method of
realizing such an enormous collecting area at an affordable price is
less clear. Looking back at Figure~1, there are indications that
some leveling out of sensitivity with time has already set in since
about 1980. This is almost certainly the result of having reached limits
in the performance to cost ratio of traditional radio telescope
technologies. We have now reached the point where system performance is
no longer limited by receiver noise, but primarily by the raw collecting
area itself. Traditional technologies have not yet made great progress
in reducing the cost of raw collecting area by orders of magnitude.
The most cost effective designs from this point of view have been the
Arecibo fixed spherical reflector and the GMRT low mass paraboloid. How
might we proceed to even greater cost-effectiveness for the unit
telescopes?

Several possible element concepts for the SKAI are illustrated in
Figure~9. At the heart of each of these concepts is a much greater
reliance than ever before on mass produced and highly integrated
receiver systems together with much more extensive digital electronics
for beam formation. In the top panel we depict one conceivable extreme
in a continuous range of possibilities. In this case the wavefront is
detected by individual active elements comparable to a wavelength in
size. Each of these is amplified, digitized and combined with the others
to form an electronically scan-able beam (or beams) with no moving
parts whatsoever. The challenge in this case lies in achieving
extremely low component and data distribution costs since literally
millions of active elements will be required.
In the center panel, some degree of field
concentration is first achieved with the use of small paraboloids before
amplification, digitization and beam formation. In this case, active
element number is reduced to some thousands and greater sky coverage at
high sensitivity is also realized, although at the expense of the mechanical
complexity of the paraboloid drive and tracking system. In the lower
panel we depict the other conceptual extreme, whereby a single large
reflector is used for each of the unit telescopes. Extensive arrays
of active elements would be employed in this concept to intercept the
focal region of the spherical primary in order to efficiently illuminate
the surface and allow multiple beams to be formed.

\begin{figure}[htb]
\caption{Possible element concepts for the SKAI. }
\end{figure}

The adaptive beam formation technology which underlies all of the
element concepts just considered is extremely attractive for a number of
reasons. Real-time beam formation with at least thousands if not
millions of active elements provides a comparable number of degrees of
freedom for tailoring the beam in a desired way. The basic properties
of high gain in some direction and low side-lobe levels elsewhere are
fairly obvious and traditional requirements. An additional possibility,
which hasn't yet been applied in radio astronomy, is that of placing
response {\it minima\ } in other desired directions, such as those of
interfering sources. In addition, the way is naturally opened to
exploit multiple observing beams on the sky to enhance the astronomical
power of the instrument many-fold. These might be used to provide
simultaneous instrumental calibration, support multiple, fully
independent observing programs or enlarge the instantaneous
field-of-view for wide-field applications. Finally, the great potential
of adaptive beam formation has led to a strong commercial interest in
this technology. This has opened the way to collaborative R\&D efforts
which are now beginning to take shape.

During the interval 1995--2000, a concerted effort at R\&D for SKAI
will be undertaken both within the NFRA and at collaborating institutes.
The various concepts depicted in Figure~9 (and potentially new ones)
will be worked out in sufficient detail to allow realistic cost
estimates to be made. Proto-typing of cost effective technologies as an
extension to the WSRT array is planned for the period 2001--2005.
Assuming the successful completion of both technical preparations and
funding arrangements, construction of the instrument is envisioned for
the period 2005--2010.

\end{document}